\documentclass[iop,apjl]{emulateapj}

\usepackage{amssymb,natbib,graphicx,floatflt,wrapfig}
\shorttitle{Mid-IR FORCAST/SOFIA Observations of M82}
\shortauthors{Nikola et al.}

\begin{document}

\title{Mid-IR FORCAST/SOFIA Observations of M82}

\author{T. Nikola\altaffilmark{1}, T.L. Herter\altaffilmark{1}, W.D. Vacca\altaffilmark{2}, J.D. Adams\altaffilmark{1}, J.M. De Buizer\altaffilmark{2}, G.E. Gull\altaffilmark{1}, C.P. Henderson\altaffilmark{1}, L.D. Keller\altaffilmark{3}, M.R. Morris\altaffilmark{4}, J. Schoenwald\altaffilmark{1}, G. Stacey\altaffilmark{1}, A. Tielens\altaffilmark{5}}

\altaffiltext{1}{Department of Astronomy, Cornell University, Ithaca, NY 14853, USA}
\altaffiltext{2}{Universities Space Research Association, NASA Ames Research Center, MS 211-3, Moffett Field, CA 94035, USA}
\altaffiltext{3}{Ithaca College, Ithaca, NY 14850, USA}
\altaffiltext{4}{Department of Physics and Astronomy, University of California, Los Angeles, CA 90095-1547, USA}
\altaffiltext{5}{Leiden Observatory, PO Box 9513, Leiden, 2300 RA, Netherlands}

\begin{abstract}
We present $75''\times75''$ size maps of M82 at 6.4 $\mu$m, 6.6 $\mu$m, 7.7 $\mu$m, 31.5 $\mu$m, and 37.1 $\mu$m with a resolution of $\sim$4$''$ that we have obtained with the mid-IR camera FORCAST on SOFIA.
We find strong emission from the inner $60''$ ($\sim$1kpc) along the major axis, with the main peak $5''$ west--southwest of the nucleus and a secondary peak $4''$ east--northeast of the nucleus.
The detailed morphology of the emission differs among the bands, which is likely due to different dust components dominating the continuum emission at short mid-IR wavelengths and long mid-IR wavelengths.
We include {\it Spitzer}-IRS and {\it Herschel}/PACS 70~$\mu$m data to fit spectral energy distribution templates at both emission peaks.
The best fitting templates have extinctions of $A_{V}=18$ and $A_{V}=9$ toward the main and secondary emission peak and we estimated a color temperature of 68~K at both peaks from the 31 $\mu$m and 37 $\mu$m measurement.
At the emission peaks the estimated dust masses are on the order of $10^{4}\ M_{\odot}$.
\end{abstract}


\section{Introduction}

Over the last few $10^{8}$~yr a series of starbursts has been triggered in M82 \citep{deGrijs.2001,Mayya.Bressan.Carrasco.2006} due to its interaction with M81 and NGC~3077 \citep[e.g.][]{Appleton.Davies.Stephenson.1981,Yun.Ho.Lo.1994,Sun.Zhou.Chen.2005,deMello.Smith.Sabbi.2008}.
The most recent starburst \citep[$\lesssim 50$~Myr, e.g.][]{Rieke.Lebofsky.Thompson.1980,Satyapal.Watson.Pipher.1997} has created a stellar cluster at the center of M82.
A ring of ionized gas that is enveloped by a molecular gas ring surrounds this cluster, and the starburst is likely fueled by gas that is funneled toward this region by a $\sim$1~kpc ($\sim$1$'$) stellar bar \citep{Telesco.Joy.Dietz.1991,Achtermann.Lacy.1995,Wills.Das.Pedlar.2000,Greve.Wills.Neininger.2002}.
These gas rings, which appear between about $10''$ and $30''$ from the nucleus, could be gas swept up by a central expanding superbubble \citep{Matsushita.Kawabe.Matsumotot.2000,Matsushita.Kawabe.Kohno.2005}.
High external gas pressure imposed on the molecular clouds probably drives the current starburst \citep{Keto.Ho.Lo.2005}.
This view is supported by the detection of warm molecular gas that is likely heated by dissipation of turbulence \citep{Panuzzo.Rangwala.Rykala.2010}.

The view to the center of M82 is obscured by patchy and high extinction \citep{Satyapal.Watson.Pipher.1995,FoersterSchreiber.Genzel.Lutz.2001}.
The mid-infrared (mid-IR) wavelength regime, where the extinction is modest, is therefore used extensively to study the center of M82 \citep[e.g.][]{Rieke.Lebofsky.Thompson.1980,Telesco.Joy.Dietz.1991,FoersterSchreiber.Genzel.Lutz.2001,Beirao.Brandl.Appleton.2008}. Individual star forming clouds that are located at the inner edge of the molecular gas ring or within the ionized gas ring have been observed in the mid-IR with large ground-based observatories that provide (sub-) arcsecond resolution \citep{Lipscy.Plavchan.2004,Gandhi.Isobe.Birkinshaw.2011}.
Unfortunately, ground-based mid-IR observations are limited to a few telluric windows at $\lesssim30\ \mu$m.

We have used the mid-IR camera FORCAST on SOFIA to study the central region of M82.
FORCAST provides relatively high spatial resolution and also covers the entire 5-40 $\mu$m wavelength range.
Emission from hot, small grains dominates the continuum at short mid-IR wavelengths while emission from warm, large grains dominates at long mid-IR wavelengths.
A total wavelength coverage is therefore important to determine the detailed contribution of each dust component to the continuum emission and to estimate the heating of the dust.

We adopt a distance  of 3.5~Mpc for M82 \citep{Dalcanton.Williams.Seth.2009}, so that $1''$ corresponds to 17.5~pc.

\section{Observation}

We observed M82 in the 6.35 $\mu$m ($\Delta \lambda = 0.14 \mu$m), 6.61 $\mu$m ($\Delta \lambda = 0.24 \mu$m), 7.71 $\mu$m ($\Delta \lambda = 0.47 \mu$m), 31.5 $\mu$m ($\Delta \lambda = 5.7 \mu$m), and 37.1 $\mu$m ($\Delta \lambda = 3.3 \mu$m) passband filters using FORCAST \citep{Adams.Herter.Gull.2010, Herter.Adams.deBuizer.2012} on SOFIA \citep{Young.Herter.deBuizer.2012}.

FORCAST is a dual-channel camera with a short-wavelength channel ($5-25~\mu$m) and a long-wavelength channel ($25-40~\mu$m), allowing for simultaneous observations at two specific wavelengths by selecting band-pass filters in filter wheels.
A dichroic beam-splitter directs the same field of view to both the short- and long-wavelength cameras.
To increase sensitivity, the beam-splitter can be either removed or replaced by a mirror during an observation.
The effective field of view of the two 256 square arrays is $3.4'\times3.2'$.
After distortion correction, the pixel size is $0.768''$.

We observed M82 on 2010 December 1 in the two long wavelength bands without dichroic at an altitude of about 43000 feet. 
We used the on-chip chop-nod observing mode with a chopper throw and nod-distance of $120''$ and $90''$, respectively, and a chop frequency of 2 Hz.
In this mode the source is alway on the detector array.
For the 37 $\mu$m and 31 $\mu$m observations we took seven and five integrations, respectively, with an on-source time of 60~s per integration.
The observations in the 37 $\mu$m band were taken at a zenith angle of about $58^{\circ}$.

On 2010 December 4, we obtained four integrations of M82 in the 6.6 $\mu$m and 7.7 $\mu$m bands and eight integrations in the 6.4 $\mu$m band from an altitude of about 42000 feet.
All bands were observed without beam-splitter, except for the 7.7~$\mu$m band observations, which were observed simultaneously with the 31 $\mu$m band.
We observed again in the on-chip chop-nod mode, but with a chopper throw and nod-distance of $90''$ and a chop frequency of 5 Hz.
All bands were observed at a zenith angle of about $53^{\circ}$.

Since M82 is bright in all bands, we spatially registered the individual chop/nod integrations using a two-dimensional Gaussian fit over the entire galaxy before co-adding the individual integrations.
These pointing-corrections are about 3 pixels and improve the noise and the final maps.
Due to the on-chip chop-nod observing mode the extent of the final images is about $75''\times75''$.
Since pointing accuracy and drifts were on the order of several arcseconds, we determined the absolute position of the M82 observation by comparing the 7.7~$\mu$m FORCAST map with the 8~$\mu$m {\it Spitzer}/IRAC map from the Spitzer archive.

The flux densities in all bands were color-corrected.
The standard stars $\beta$Gem, $\beta$UMi, and $\mu$UMa were used for absolute flux density calibration and to estimate the beam size.
Based on the uncertainty of flat fielding and the water vapor burden, as well as the variance of the flux density measurements, the uncertainty of the intrinsic flux density of the standard stars, and the fact that $\beta$Gem and $\beta$UMi are variable stars, we estimate a $3\sigma$ uncertainty of $\sim$20\% for the absolute calibration \citep{Herter.Vacca.Adams.2012,Herter.Adams.deBuizer.2012}.
Due to turbulent airflow across the telescope and pointing instabilities the observations are not diffraction limited.
For the M82 observations, we estimate a beam size of about $4''\pm0.5''$ for all the bands.

The original maps of the 6.4 $\mu$m, 6.6 $\mu$m, and 7.7 $\mu$m continuum had a lower signal-to-noise than the two long-wavelength maps and we therefore smoothed them with a 2 pixel FWHM Gaussian kernel.
The final flux densities are presented in Table~\ref{tab1}.

We include {\it Herschel}/PACS 70~$\mu$m photometric data in our spectral energy distribution (SED) analysis below.
Since processed and published PACS 70~$\mu$m photometric data were unavailable, we used level 2.5 data product from the Herschel archive.
The observations were carried out in scan-map mode at medium speed ($20''$~s$^{-1}$) resulting in a smeared beam of about $5.46''\times5.76''$ (PACS Observer Manual).
We extracted background-subtracted flux densities within a $6.8''$ aperture ($\sim3.5$ PACS pixels) at three positions (see Table~\ref{tab1} for positions).
The flux densities are $115\pm11$~Jy (main peak), $98\pm10$~Jy (secondary peak), and $116\pm11$~Jy (ridge).
Within a radius of $35''$, similar to the entire region observed with FORCAST, the 70~$\mu$m flux density is $1644\pm41$ Jy.

\section{Results}

\subsection{Morphology}\label{sec:morph}

\begin{figure}[ht]
\centering
\includegraphics[width=0.40\textwidth]{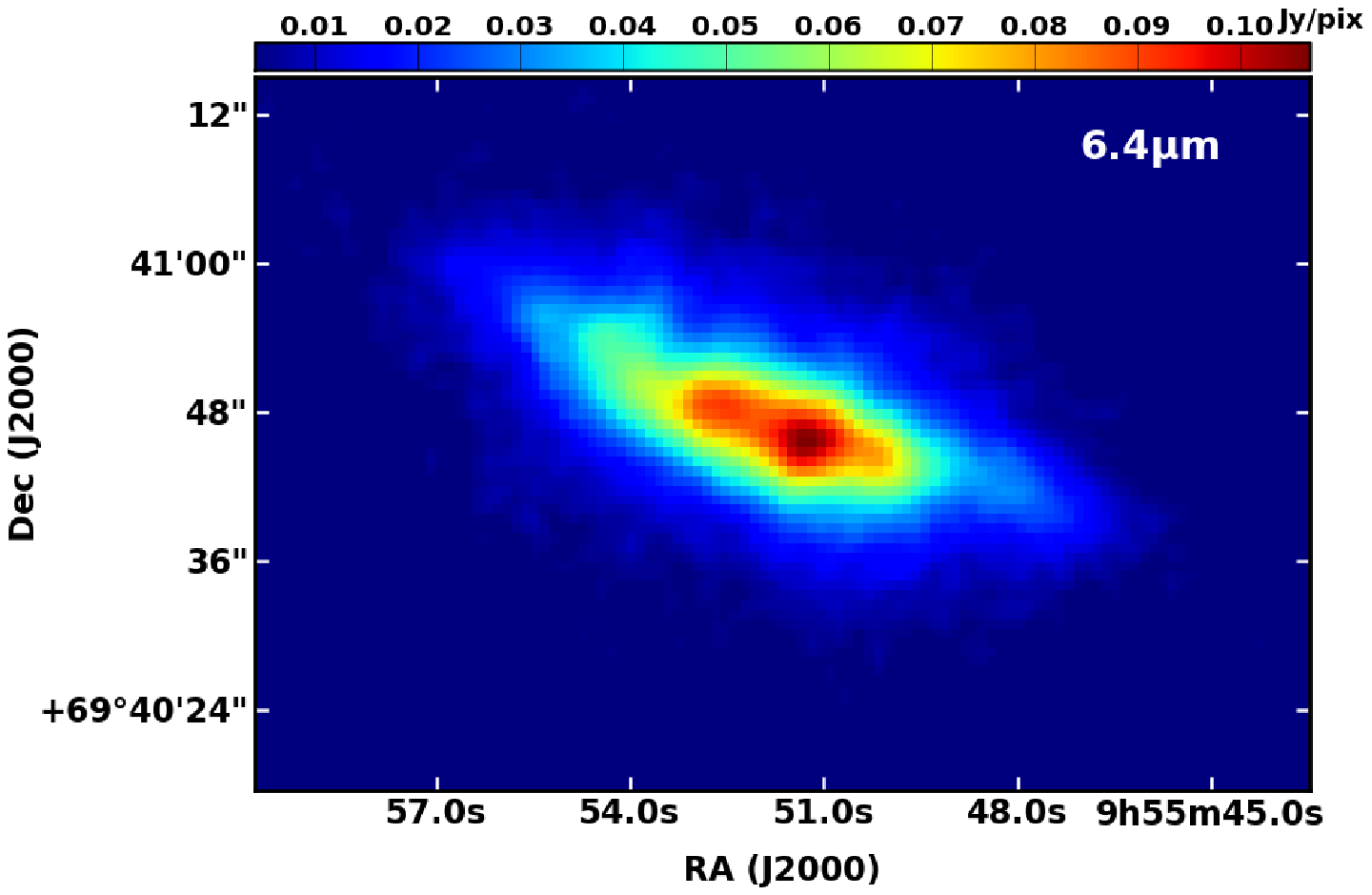}
\hspace{0.5cm}
\includegraphics[width=0.40\textwidth]{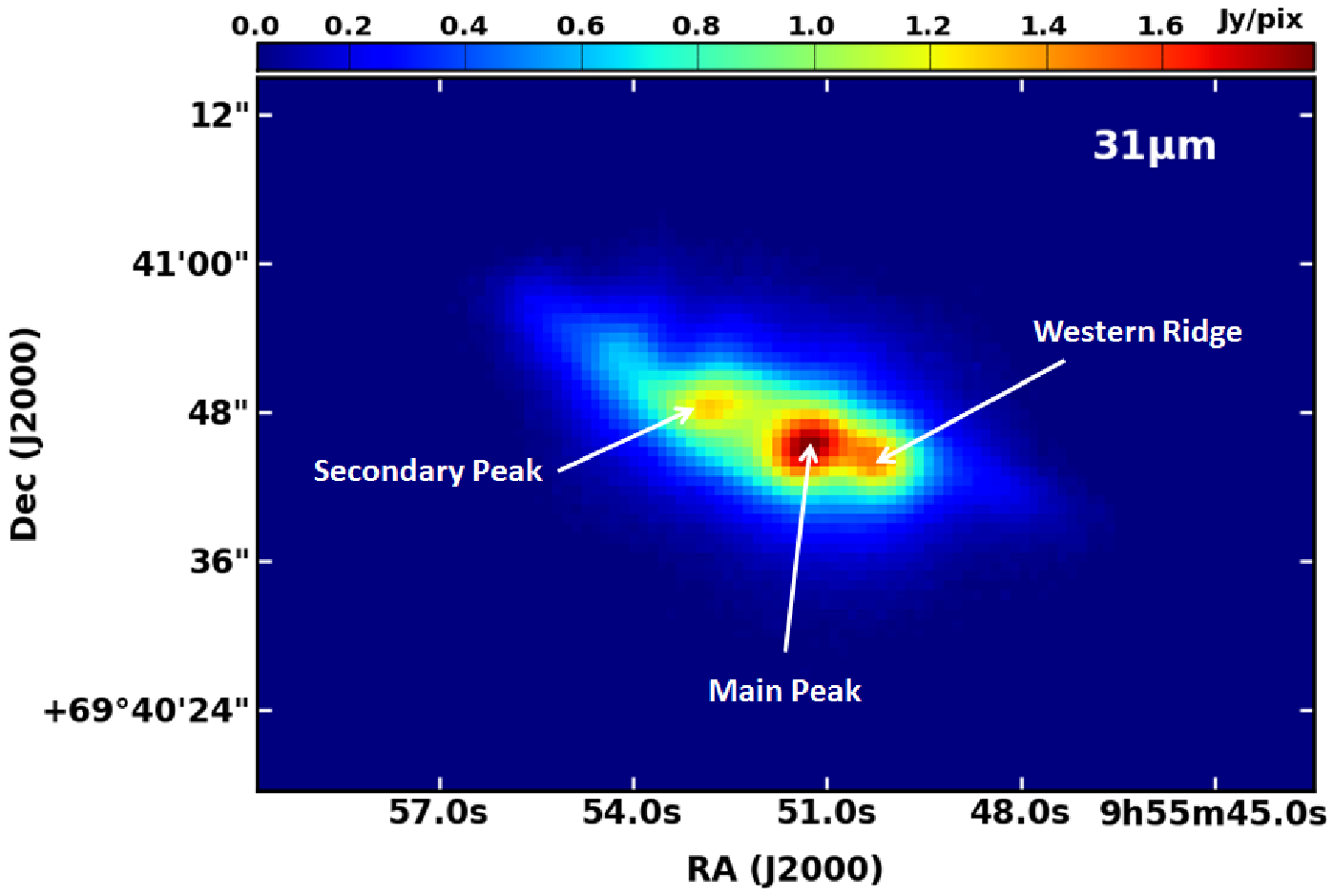} \\
\includegraphics[width=0.40\textwidth]{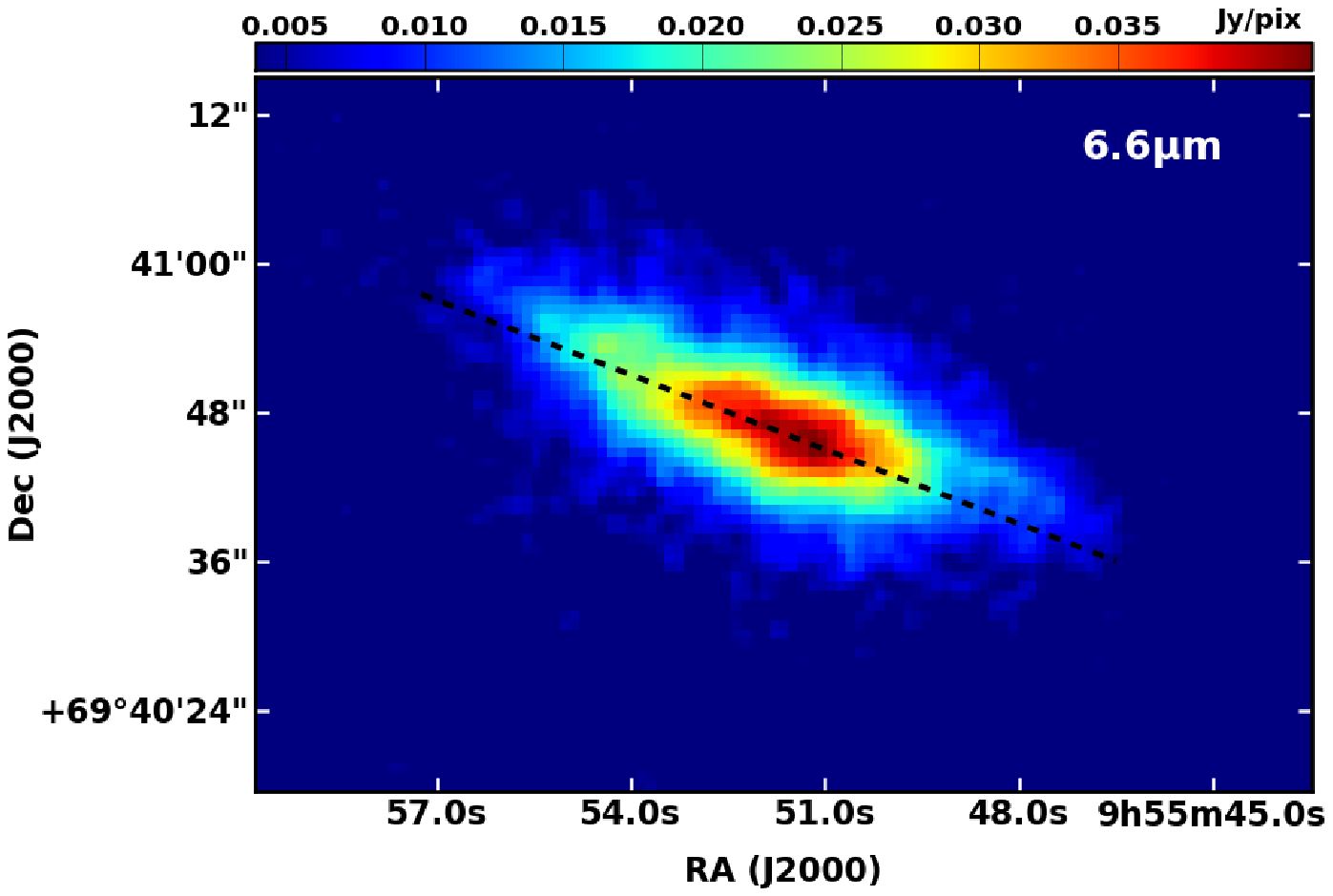}
\hspace{0.5cm}
\includegraphics[width=0.40\textwidth]{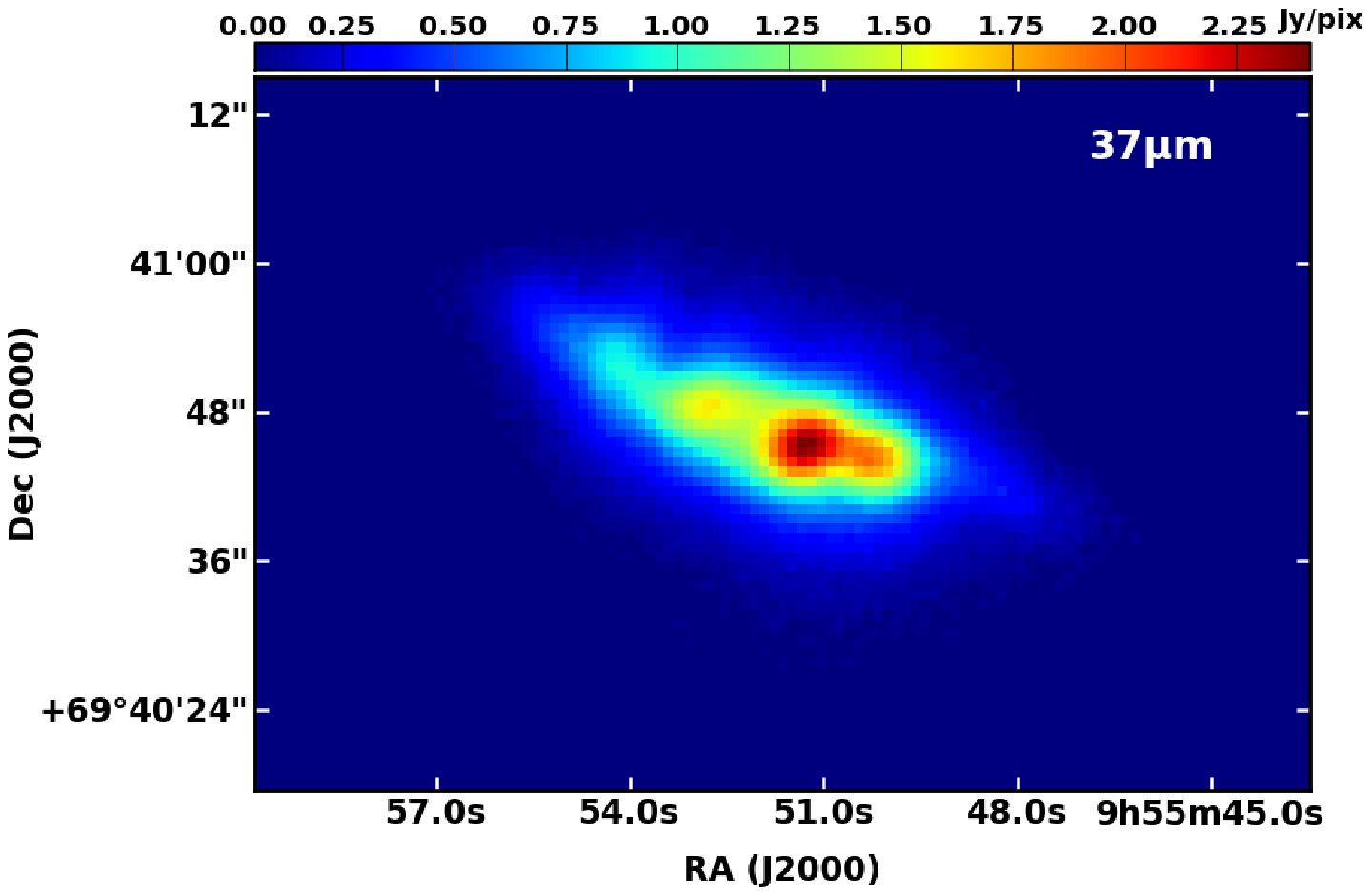} \\
\includegraphics[width=0.40\textwidth]{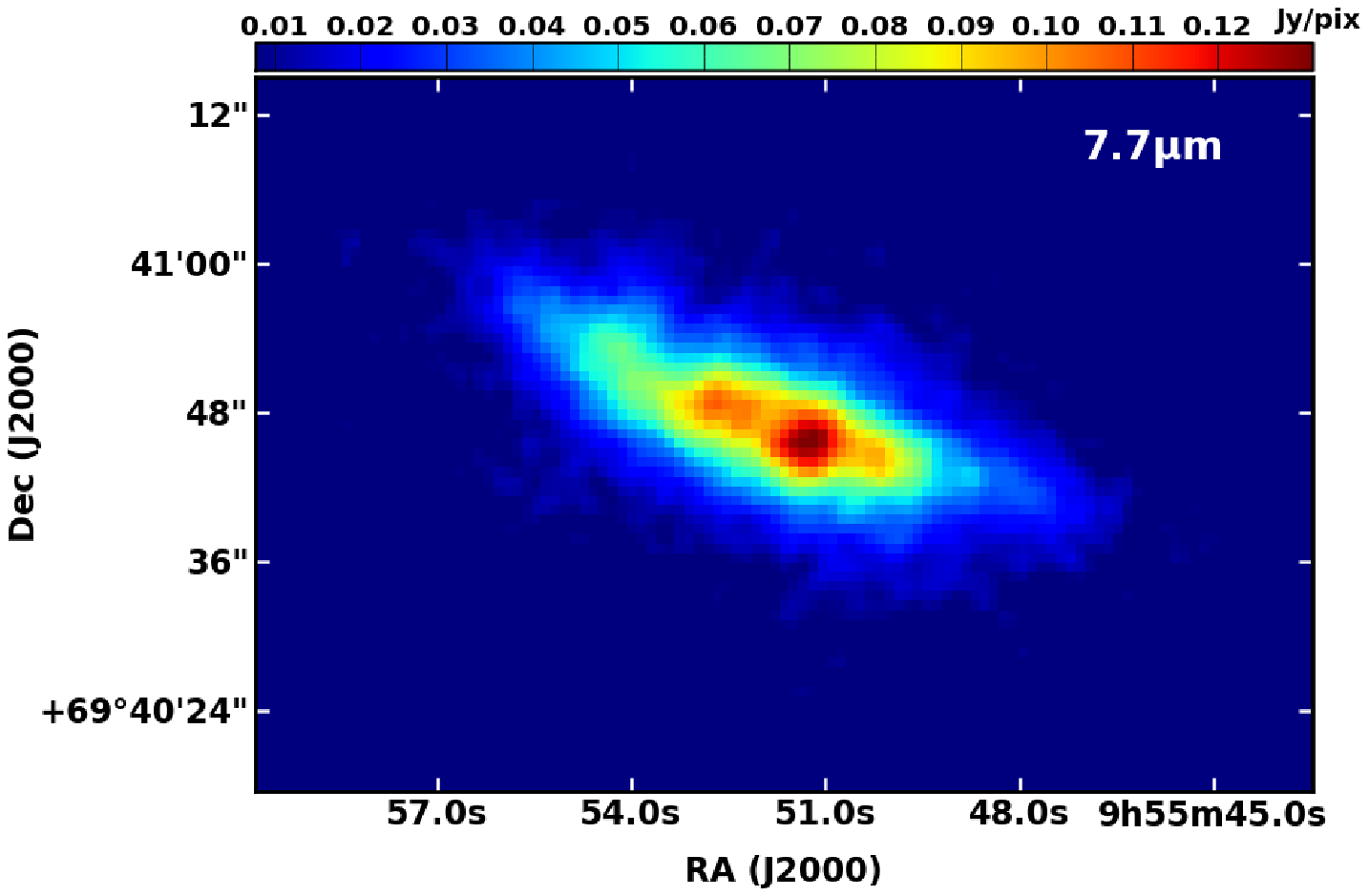}
\hspace{0.5cm}
\includegraphics[width=0.40\textwidth]{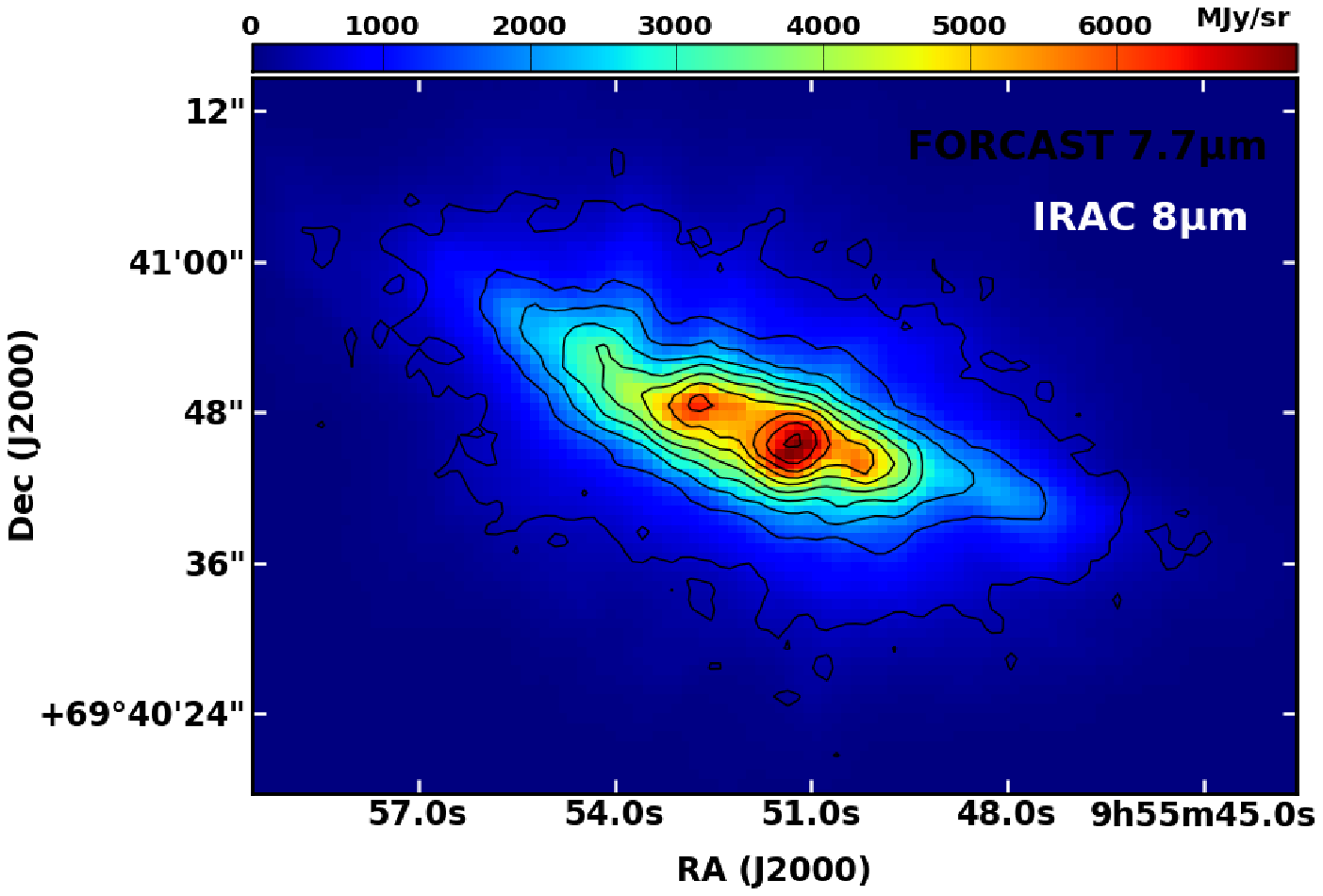}
\caption{Maps of M82 in the FORCAST bands in units of Jy pixel$^{-1}$ at 6.4 $\mu$m (top left),  6.6 $\mu$m (mid left),  7.7 $\mu$m (bottom left),  31.5 $\mu$m (top right),  and 37.1 $\mu$m (mid right). Bottom right: IRAC band 4 (8 $\mu$m) map (in MJy sr$^{-1}$) overplotted with FORCAST 7.7 $\mu$m contours . The color scale is linear and starts at the $3\sigma$ level of the statistical background noise (0.009 Jy at 6.4 and 6.6 $\mu$m, 0.018 Jy at 7.7 $\mu$m, 0.042 Jy at 31.5 $\mu$m, and 0.051 Jy at 37.1 $\mu$m). The dashed line in the 6.6 $\mu$m map (mid left) indicates the position of the profiles shown in Figure~\ref{fig:m82_profile}.}
\label{fig:m82_fcast}
\end{figure}

Figure~\ref{fig:m82_fcast} shows the distribution of the mid-IR emission in each of the observed bands.
The 6.4 $\mu$m, 6.6 $\mu$m, and 7.7 $\mu$m emission is slightly more extended than the 31 $\mu$m and 37 $\mu$m emission.
Also, the contrast in the morphology is higher in the 31 $\mu$m and 37 $\mu$m emission than in the short-wavelength bands, with the peaks being more pronounced.
The main peak of all mid-IR distributions is $4.5''$ west--southwest from the dynamical center.
Strong emission extends further to the west--southwest and ``peaks'' roughly $9''$ from the center (``western ridge''). 
A secondary peak is visible $4''$ east--northeast of the center (except at 6.6 $\mu$m).
In the 7.7 $\mu$m band, the emission at the secondary peak appears as two separate components.
The positions of the main and secondary peaks coincide roughly with the position of the ionized ring or the inner edge of the molecular ring.

The 7.7~$\mu$m FORCAST map is in excellent agreement with the {\it Spitzer}/IRAC 8~$\mu$m map  (Figure~\ref{fig:m82_fcast}).
Also, the FORCAST 31 $\mu$m map agrees very well with the 30 $\mu$m map obtained by \citet{Telesco.Joy.Dietz.1991}  using the NASA Infrared Telescope Facility (IRTF), and the flux densities at the main and secondary peaks agree within the calibration uncertainty.

\begin{figure}[ht]
\centering
\includegraphics[width=0.90\textwidth]{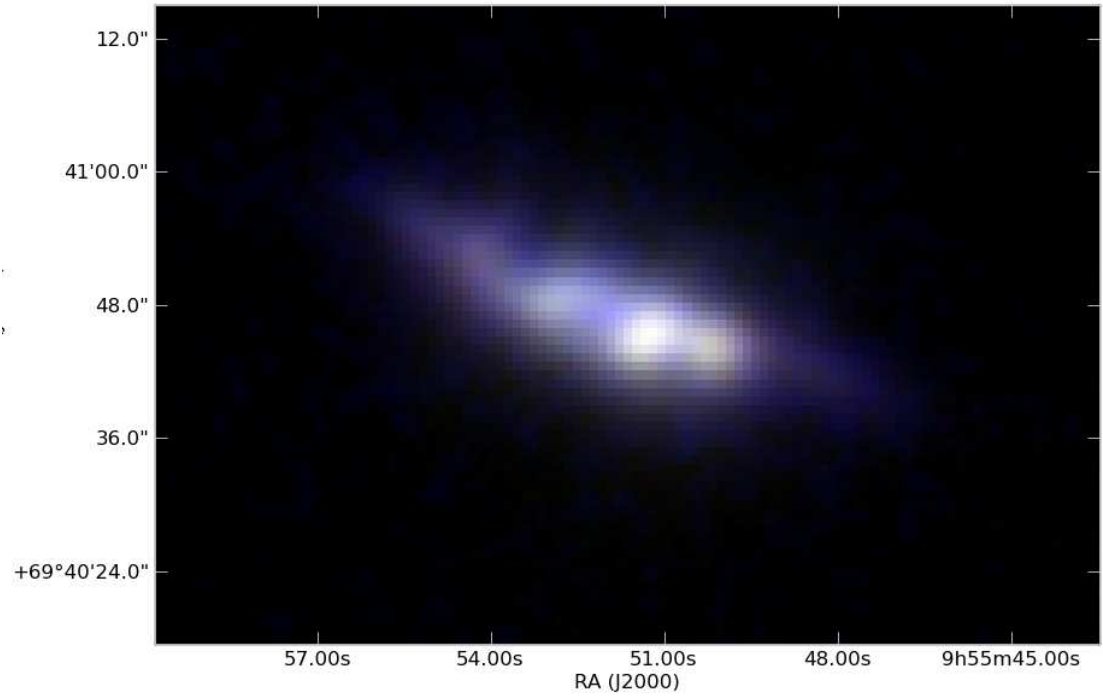}
\caption{Three color image of M82, with the FORCAST 6.6 $\mu$m as red, 31.5 $\mu$m as green, and 37.1 $\mu$m as blue. All bands are linearly scaled, starting from 3$\sigma$ of the statistical background noise, before combined.}
\label{fig:m82_3color}
\end{figure}

In Figure~\ref{fig:m82_3color} we show a three-color image of M82, where we have combined the 6.6 $\mu$m (blue), 31 $\mu$m (green), and 37 $\mu$m (red) bands.
All colors are scaled linearly with flux density and start at the 3$\sigma$ level of the statistical background noise in each band.
The image shows that the 6.6 $\mu$m is more extended and that the secondary peak and western ridge are more pronounced in the two long-wavelength bands.

\begin{figure}[ht]
\centering
\includegraphics[width=0.40\textwidth]{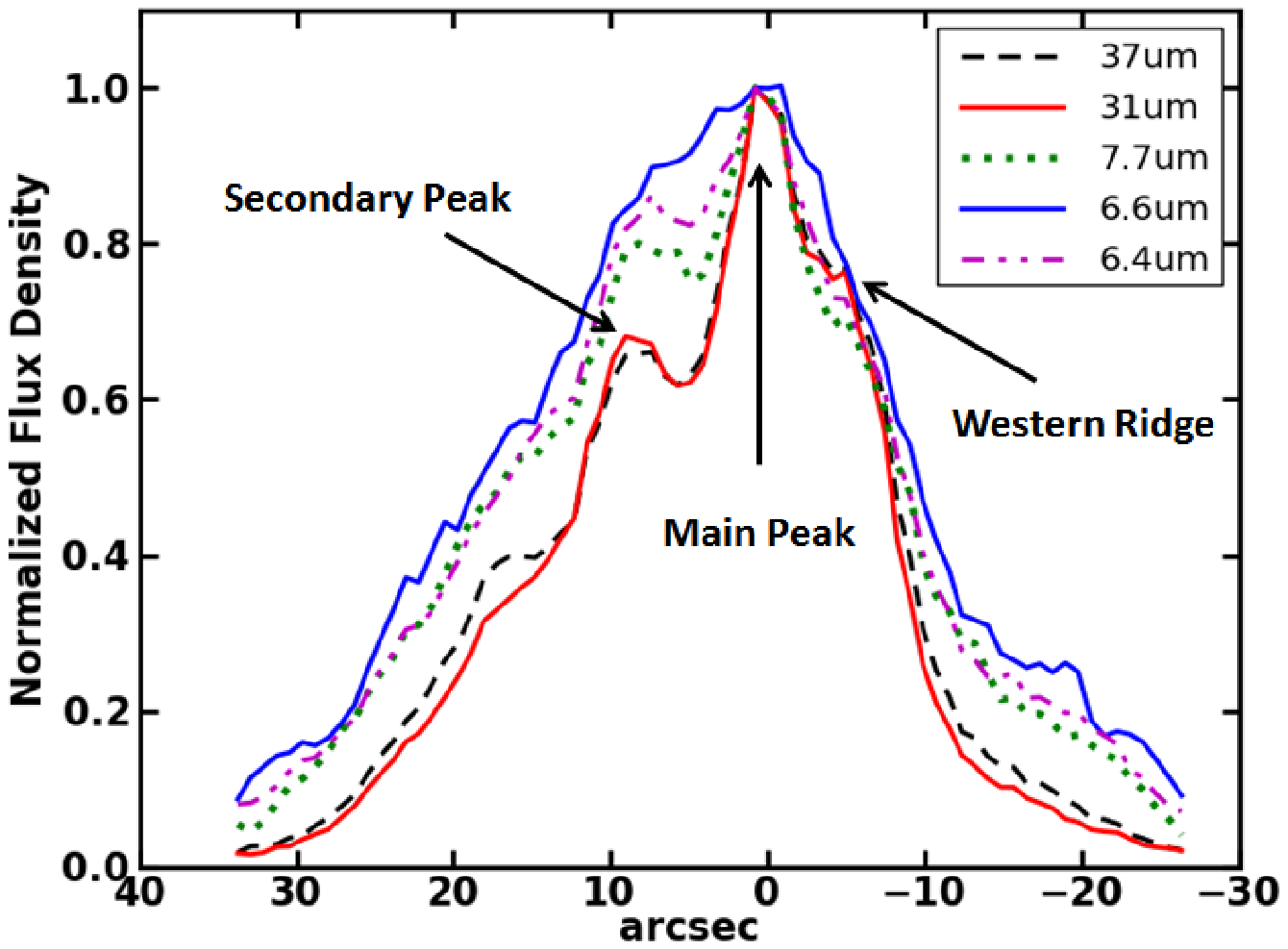}
\hspace{0.5cm}
\includegraphics[width=0.40\textwidth]{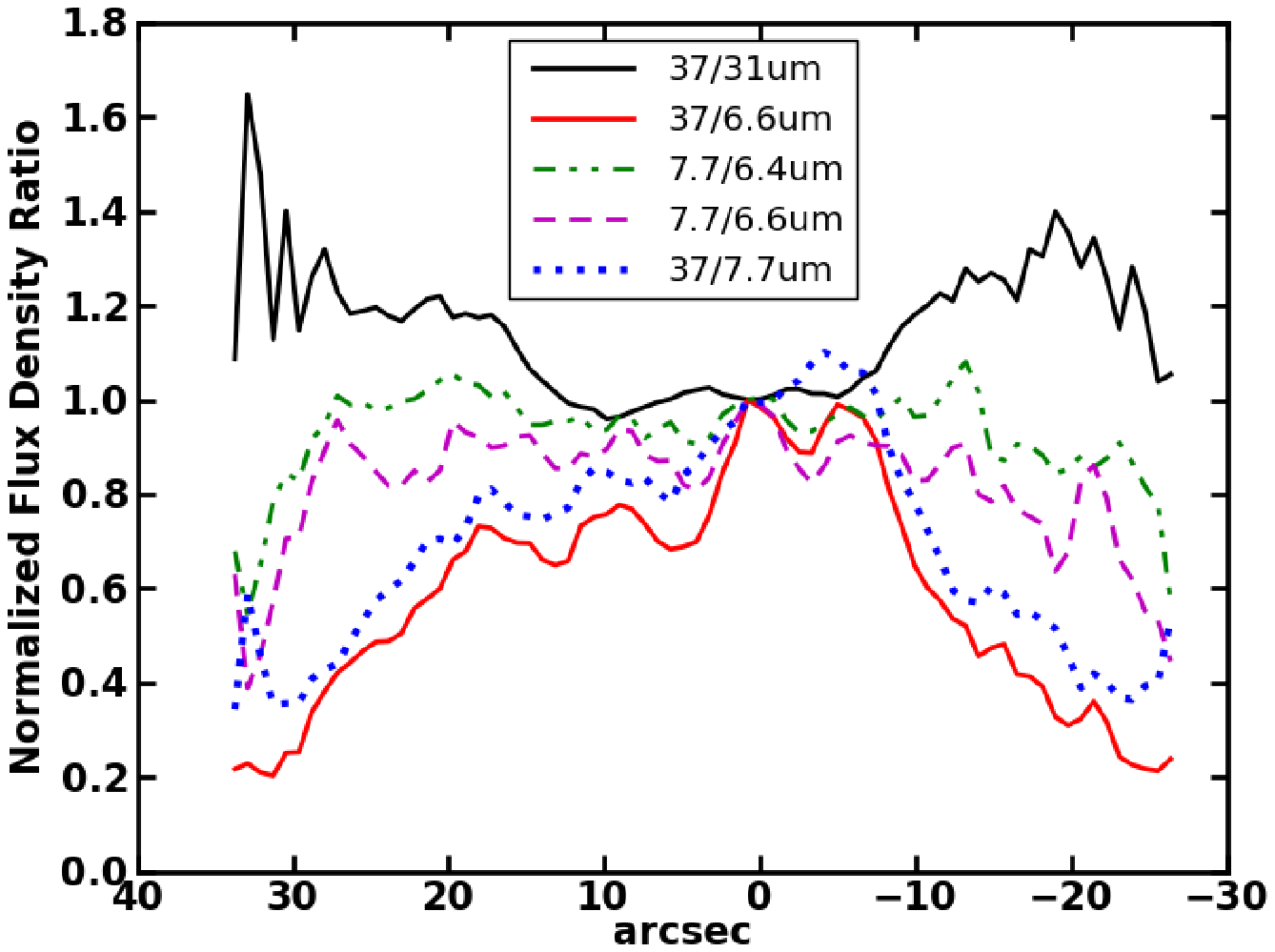}
\caption{Flux densities (left) and flux density ratios (right), normalized to the value at the main peak position, along the major axis of M82. The reference position is the main peak, distances are in arcsec, and positive distance is toward the northeast. Flux densities are summed over $1\times5$ pixels perpendicular to the major axis.}
\label{fig:m82_profile}
\end{figure}

Figure~\ref{fig:m82_profile} shows the emission profiles along a $60''$ cut following the major axis of M82.
The length and position of the cut is shown in Figure~\ref{fig:m82_fcast}.
The profiles are in steps of 1 pixel and the flux densities are summed over $1\times5$ pixel strips perpendicular to the major axis and normalized to the emission peak.
They emphasize that in the 31 $\mu$m and 37 $\mu$m bands the ratio between the main and secondary peak is much stronger than in the three short mid-IR bands.
In fact, the 6.6 $\mu$m continuum only shows a gradient in the flux density from the main peak toward the position of the secondary peak.
The emission from the ``western ridge'' is very pronounced in the 31 $\mu$m and 37 $\mu$m bands, less in the 7.7~$\mu$m and 6.4~$\mu$m maps, and barely noticeable in the 6.6 $\mu$m map.
The 31 $\mu$m and 37 $\mu$m emission show a peak at this position that is stronger than the secondary peak.

Figure~\ref{fig:m82_profile} also shows the flux density ratios along the major axis of M82.
The profile of the 37$\mu$m/6.6$\mu$m and 37$\mu$m/7.7$\mu$m flux density ratios basically follows the profile of the 37 $\mu$m flux density.
In contrast, the flux density ratios of the 7.7$\mu$m/6.6$\mu$m and 7.7$\mu$m/6.4$\mu$m bands are fairly uniform along the major axis, with larger variations only at the edge of the profile, where the signal-to-noise becomes small.
The profile of the 37$\mu$m/31$\mu$m flux density ratio differs from all others. 
It is basically flat within $\pm10''$ of the center of M82 and then increases on either side, while the flux density ratios of all other bands either decrease or stay constant with distance from the center.
This indicates a higher color temperature of the large grains in the inner region than in the outer region.
Also, the difference of the profiles between the three short mid-IR bands and two long mid-IR bands indicate that they trace different dust components with spatially distinct excitation conditions.

\subsection{Mid-IR Spectral Energy Distribution}

\begin{figure}[ht]
\centering
\includegraphics[width=0.40\textwidth]{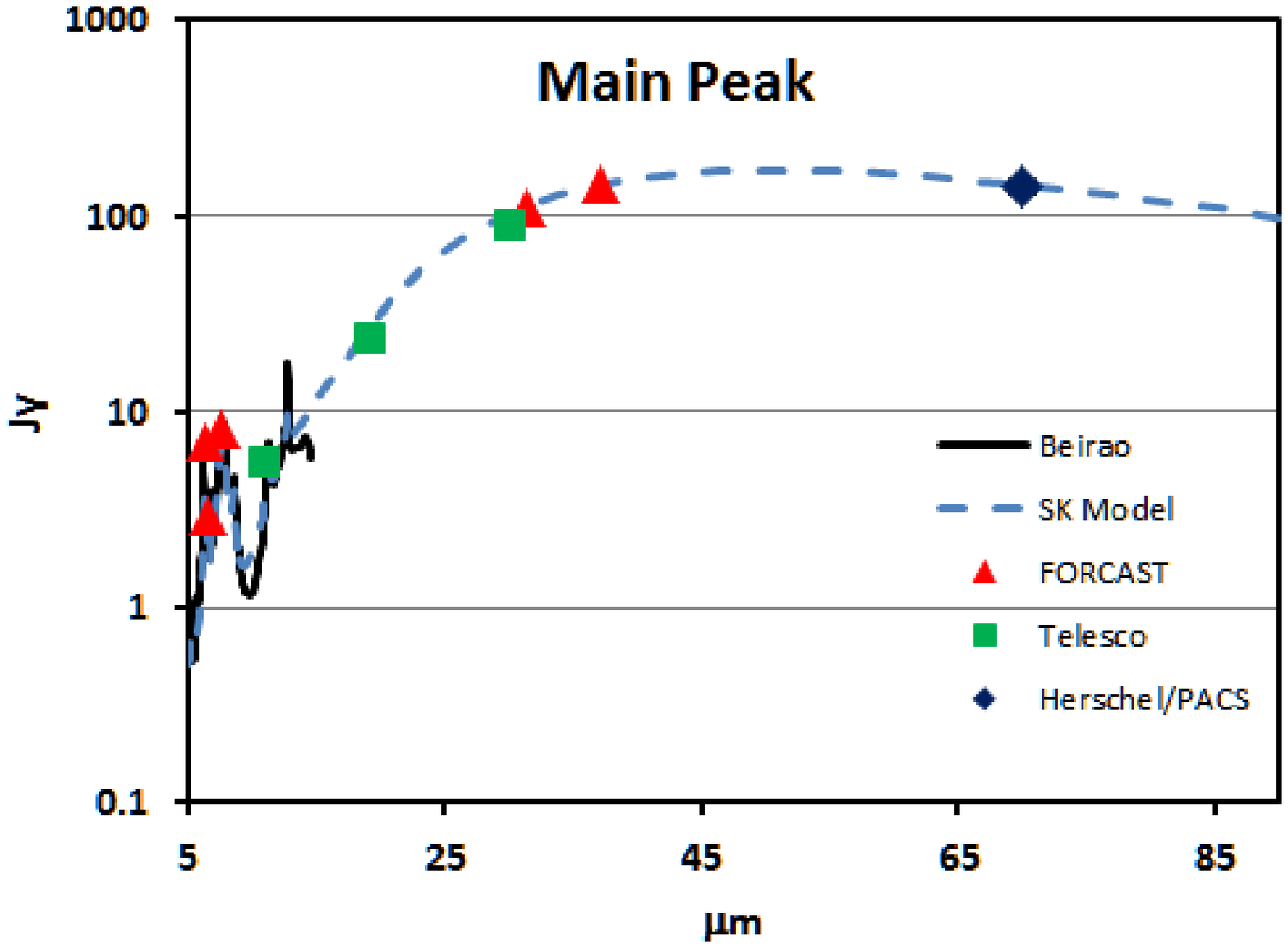}
\hspace{0.5cm}
\includegraphics[width=0.40\textwidth]{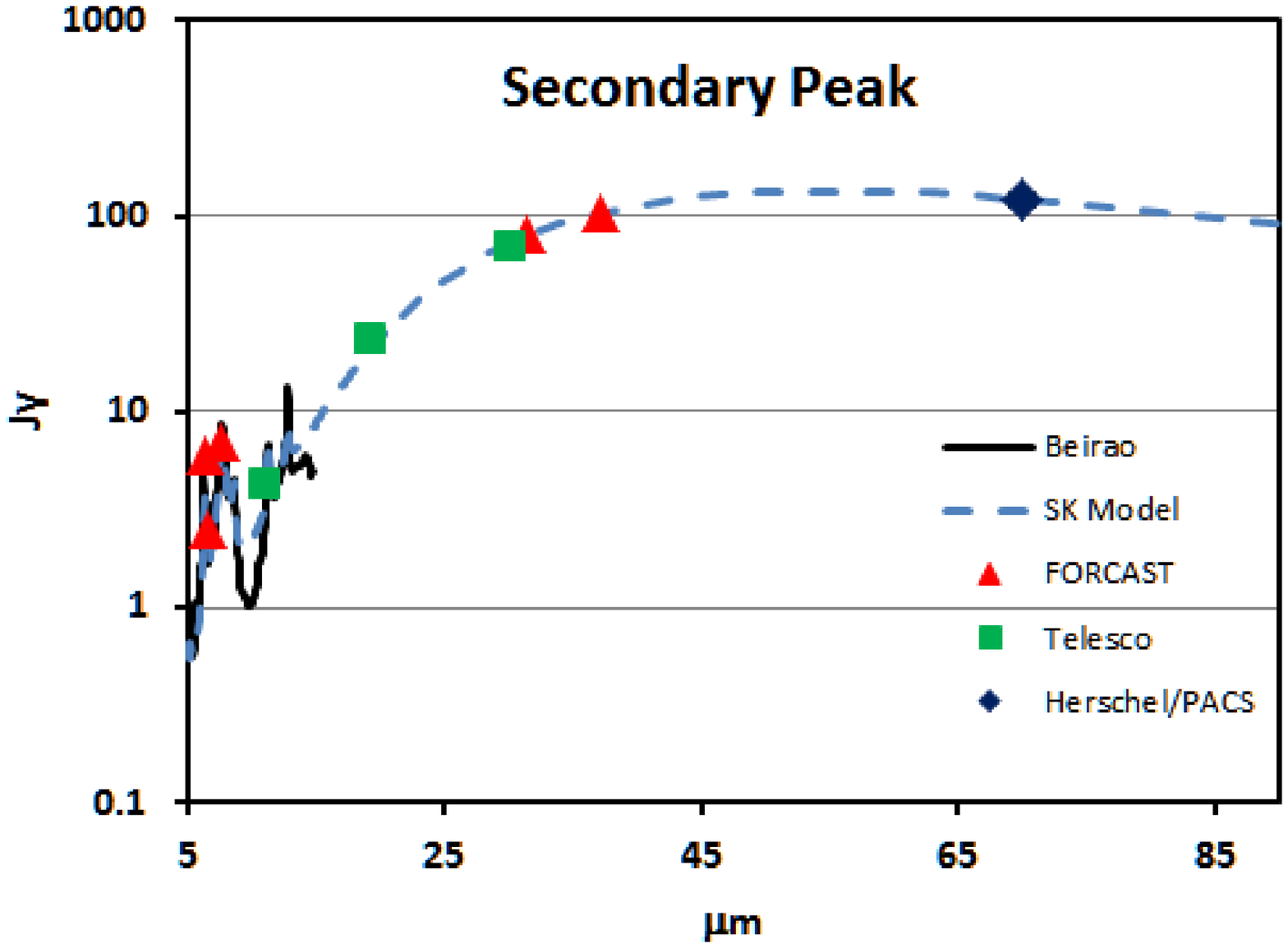}
\caption{Mid-IR SED of the main peak (left) and secondary peak (right). The solid (black) line is the low-resolution {\it Spitzer}/IRS spectrum \citep{Beirao.Brandl.Appleton.2008}, filled triangles (red) are the FORCAST observations, filled diamond (blue) is the {\it Herschel}/PACS 70 $\mu$m observation, filled squares (green) are IRTF  observations \citep{Telesco.Joy.Dietz.1991}, multiplied by a factor of two (see the text), dashed line (blue) is the Siebenmorgen \& Kruegel SED model \cite{Siebenmorgen.Kruegel.2007}. The error bars are smaller than the symbols.}
\label{fig:m82_seds}
\end{figure}

\citet{Siebenmorgen.Kruegel.2007}\footnote{www.eso.org/\symbol{126}rsiebenm/sb\_models/} have created template SEDs for starbursts as a function of the total luminosity ($L({\rm tot})$), optical extinction, contribution to the total luminosity by OB stars, and radius of the nuclear starburst, covering the wavelength range 0.03-2000 $\mu$m.
We use the FORCAST and PACS 70~$\mu$m data points as well as low-resolution {\it Spitzer}-IRS spectra between 5.3~$\mu$m to 12.8~$\mu$m from region 2 and 3 of \citet{Beirao.Brandl.Appleton.2008} to find the SED template that fits the data in the two mid-IR emission peaks within a $6.8''\times6.8''$ ($\sim 115\times115$ pc) area, corresponding to $\sim$$9\times9$ FORCAST pixels.
The lower size limit of the area is given by the {\it Spitzer} observation. 
Since this area is much smaller than the smallest nuclear radius in the SED template grid (350 pc), we constrained our search algorithm to that template series and allow it to fit an additional scaling factor.
Figure~\ref{fig:m82_seds} shows the data points and the best-fitting SED template at the main and secondary peaks.
It includes three data points from \citet{Telesco.Joy.Dietz.1991}, multiplied by a factor of two, which corresponds to the flux density ratio within a $6.8''$ and $4.5''$ aperture in a FORCAST band at the peaks.
The flux densities at 70~$\mu$m and 37 $\mu$m at the main and secondary peaks are very similar and the 70 $\mu$m flux density is thus crucial to constrain the SED fit.

For the main mid-IR emission peak, the best-fitting SED template has a total luminosity of $L_{\rm MP}({\rm tot}) = 10^{11.2}~L_{\odot}$, an OB luminosity fraction of 40\%, an extinction of $A_{V} = 18$, and a hydrogen density of $n = 5\times10^{3}$ cm$^{-3}$ in the hot spots.
The template requires an additional scaling factor of 0.041, which is of the order of the ratio of observed to the template area, and lowers the total luminosity to $L_{\rm MP}({\rm tot}) = 6.7\times10^{9}~L_{\odot}$.
The best fitting model for the secondary emission peak has a total luminosity of $L_{\rm SP}({\rm tot}) = 10^{10.7}~L_{\odot}$, an OB luminosity fraction of 40\%, an extinction of $A_{V} = 9$, a hydrogen density of $n = 5\times10^{3}$ cm$^{-3}$, and requires an additional scaling factor of 0.11, which reduces the total luminosity to $L_{\rm SP}({\rm tot}) = 5.7\times10^{9}~L_{\odot}$.

Due to limited resolution of the SED template parameters we estimate the derived extinction to be within a factor of two.
Both values are lower than the extinction reported by \citet{FoersterSchreiber.Genzel.Lutz.2001} for the entire central region ($A_{\rm V} = 52 \pm 17$).
This discrepancy could be due to the patchy extinction in the central region \citep{Satyapal.Watson.Pipher.1995,Lipscy.Plavchan.2004,Gandhi.Isobe.Birkinshaw.2011}.

\subsubsection{Extinction}

The extinction laws between 3 and 10 $\mu$m toward the center of M82 and the Galactic center (GC) are similar \citep{FoersterSchreiber.Genzel.Lutz.2001}, and for the GC it is best modeled with $R_{V} = 5.5$ \citep{Draine.2003}.
To estimate the extinction and emissivity at the FORCAST bands we therefore use the GC extinction law as modeled by \citet{Li.Draine.2001} and \citet{Weingartner.Draine.2001a}\footnote{www.astro.princeton.edu/\symbol{126}draine/dust/dustmix.html} with $R_{V} = 5.5$, a gas-to-dust mass ratio of 105, a dust grain mass per hydrogen atom of $m_{d} = 2.2\times10^{-26}$ g/H, $N({\rm H})/A_{V} = 1.37\times10^{21}$ cm$^{-2} \ {\rm mag}^{-1}$ (with $N({\rm H}) =N({\rm HI})+N({\rm H}_{2})$), and an emissivity index of $\beta = 1.79$ for the 31 $\mu$m and 37 $\mu$m bands.
The mass absorption coefficient and albedo at 31.62 and 37.15~$\mu$m are $\kappa = 385.2$ cm$^{2}$g$^{-1}$ and $0.0007$, and $\kappa = 288.8$ cm$^{2}$g$^{-1}$ and $0.0006$, respectively.
Using the relation $\kappa_{\rm abs}(\lambda) = 0.4\ \ln 10 \ (1 - {\rm albedo}) \times [ A(\lambda)/N_{\rm H}]/m_{\rm dust}$ cm$^{2}$g$^{-1}$ \citep{Li.Draine.2001} leads to $A(37)/A_{V} = 9.46\times10^{-3}$ and $A(31)/A_{V} = 1.26\times10^{-2}$, resulting in $\tau(37\mu{\rm m}) = 0.16$ and $\tau(31\mu{\rm m}) = 0.21$ at the main emission peak, respectively.
At the secondary emission peak the opacities are $\tau(37\mu{\rm m}) = 0.08$ and $\tau(31\mu{\rm m}) = 0.11$.

\subsubsection{Dust Mass}

We use three different methods to estimate the dust masses at the main and secondary emission peaks within $6.8''\times6.8''$.

Applying the extinction laws from the previous section we estimate the gas mass using $M_{\rm g} = \mu \times m_{\rm H} \times N({\rm H}) \times {\rm Area}$, where $\mu=1.4$ is the mean atomic mass per hydrogen, $m_{\rm H}$ is the mass of atomic hydrogen, and $N({\rm H})$ is the total hydrogen column density.
We obtain a total gas mass of $M_{\rm g}({\rm MP}) = 3.8\times10^{6} \ M_{\odot}$  at the main peak and $M_{\rm g}({\rm SP}) = 1.9\times10^{6} \ M_{\odot}$ at the secondary peak.
Using a gas-to-dust ratio of 105 this gives dust masses of $M_{\rm d}({\rm MP}) = 3.6\times10^{4} \ M_{\odot}$ and $M_{\rm d}({\rm SP}) = 1.8\times10^{4} \ M_{\odot}$ at the main and secondary peak, respectively.
Due to the uncertainty of the extinction, the dust masses can vary by a factor of two.

The dust mass can also be estimated by $M_{\rm d} = 1 / \kappa_{\rm abs} \times F_{\nu}(\lambda) \times D^{2} / B(\lambda, T)$, where $\kappa_{\rm abs}$ is the mass absorption coefficient, $F_{\nu}$ is the flux density, $D$ is the distance to M82, and $B(\lambda, T)$ is the Planck function at wavelength $\lambda$ and temperature $T$.
Using the previously estimated opacities for 31~$\mu$m and 37~$\mu$m and a modified blackbody function we estimate a color temperature of $68\pm10$~K at both emission peaks.
The resulting dust masses are $1.15\times10^{4} \ M_{\odot}$ and $7.9\times10^{3} \ M_{\odot}$ at the main and secondary peaks, respectively.
The uncertainty of the masses is within a factor of four.

\citet{Sanders.Scoville.Soifer.1991} estimate the dust mass using the relation $M_{\rm d} = (L({\rm FIR}) / 10^{8}\ L_{\odot}) \times (40 {\rm K}/ T_{\rm d})^{5} \ 10^{4}\ M_{\odot}$, where $T_{\rm d}$ is the dust temperature and $L({\rm FIR})$ is the far-IR luminosity in the range between 40 and 500 $\mu$m.
From the best fitting model SED we determine a $L_{\rm MP}({\rm FIR}) = 3.08\times10^{9}L_{\odot}$ at the main peak and $L_{\rm SP}({\rm FIR}) = 2.52\times10^{9}L_{\odot}$ at the secondary peak.
Using the color temperatures derived above we obtain dust masses of $M_{\rm d}({\rm MP})=2.2\times10^{4}M_{\odot}$ and 
$M_{\rm d}({\rm SP})=1.8\times10^{4}M_{\odot}$ at the main and secondary peak.
These dust masses are good to a factor of three.

High resolution CO measurements with a $4.2''$ aperture suggest molecular hydrogen column densities of $N({\rm H}_{2}) \approx (4 -- 10) \times 10^{22}$ cm$^{-2}$ at our main peak and $N({\rm H}_{2}) \approx 4 \times 10^{22}$ cm$^{-2}$ at our secondary peak \citep{Weiss.Neininger.Huettenmeister.2001}.
So, the enclosed gas masses, including He and other heavy elements, are $(1.7--4.2) \times 10^{6} \ M_{\odot}$.
Assuming a gas-to-dust ratio of 105 this yields dust masses between $(1.6--4) \times 10^{4} \ M_{\odot}$, in agreement with our calculations above.

\section{Summary}

We have presented the first results of mid-IR observations of M82 obtained with FORCAST on SOFIA.
M82 was observed in the 6.4 $\mu$m, 6.6 $\mu$m, 7.7 $\mu$m, 31.5 $\mu$m, and 37.1 $\mu$m bands.
The observations cover a $75''\times75''$ region.
All bands show a strong peak located $4.5''$ west--southwest of the kinematic center of M82.
A secondary peak $4''$ east--northeast of the nucleus is seen in the 6.4 $\mu$m, 7.7 $\mu$m, 31.5 $\mu$m, and 37.1 $\mu$m bands, but not in the 6.6 $\mu$m continuum.
The profiles of the flux density ratios over $60''$ along the major axis indicates that the emission at the three short mid-IR bands is dominated by a different dust component than the emission at the two long mid-IR bands. 
We fitted SED templates to the FORCAST data combined with PACS 70 $\mu$m and low-resolution {\it Spitzer}-IRS spectra and estimated extinctions of $A_{V}=18$ and $A_{V}=9$ toward the main and secondary peaks and a dust color temperature of 68~K in both peaks.
The dust masses at the locations of the emission peaks are on the order $10^{4}\ M_{\odot}$.

\acknowledgments
We thank SOFIA telescope engineering and operations team, as well as the SOFIA flight crews and mission operations team for their support.
This work is based on observations made with FORCAST on the NASA/DLR Stratospheric Observatory for Infrared Astronomy (SOFIA). SOFIA science mission operations are conducted jointly by the Universities Space Research Association, Inc. (USRA), under NASA contract NAS2-97001, and the Deutsches SOFIA Institut (DSI) under DLR contract 50 OK 0901. Financial support for FORCAST was provided to Cornell by NASA through award 8500-98-014 issued by USRA. 
This work is based in part on observations made with the Spitzer Space Telescope, obtained from the NASA/ IPAC Infrared Science Archive, both of which are operated by the Jet Propulsion Laboratory, California Institute of Technology under a contract with the National Aeronautics and Space Administration.
This research has made use of NASA’s Astrophysics Data System Abstract Service.

\clearpage

\begin{deluxetable}{c c c c c c c c}
\tabletypesize{\footnotesize}
\tablecaption{M82 FORCAST Flux Densities \label{tab1}}
\tablehead{
\colhead{Band}  &  \colhead{$S_{\rm peak}$$^{\rm a}$} & \colhead{$S_{\rm mp}$$^{\rm b,c}$}  & \colhead{$S_{\rm sp}$$^{\rm b,d}$}  & \colhead{$S_{\rm wr}$$^{\rm b,e}$}  & \colhead{$S_{\rm mp}$$^{\rm f,c}$}  & \colhead{$S_{\rm sp}$$^{\rm f,d}$}  &  \colhead{$S$(Total Map)$^{\rm g}$}  \\
\colhead{($\mu$m)}  &  \colhead{(Jy pixel$^{-1}$)}  &  \colhead{(Jy)}   &  \colhead{(Jy)}   &  \colhead{(Jy)}   &  \colhead{(Jy)}   &  \colhead{(Jy)}   &  \colhead{(Jy)} 
}
\startdata
6.4  &  $0.112 \pm 0.007$   &  $3.42 \pm 0.23$  &  $2.96 \pm 0.20$  &  $2.75 \pm 0.18$  & $6.93 \pm 0.46$ & $6.02 \pm 0.40$ &  $68 \pm 5$ \\
6.6  &  $0.047 \pm 0.003$   &  $1.40 \pm 0.09$  &  $1.21 \pm 0.08$  &  $1.17 \pm 0.08$  & $2.90 \pm 0.19$ & $2.52 \pm 0.17$ &  $32 \pm 2$ \\
7.7  &  $0.141 \pm 0.009$   &  $4.12 \pm 0.28$  &  $3.41 \pm 0.23$  &  $3.20 \pm 0.21$  & $8.16 \pm 0.54$ & $6.94 \pm 0.46$ &  $75 \pm 5$ \\
31.5  &  $1.86 \pm 0.12$  &  $57.1 \pm 3.8$   &  $40.6 \pm 2.7$   &  $46.5 \pm 3.1$   & $110.0 \pm 7.3$ & $79.7 \pm 5.3$ &  $676 \pm 45$ \\
37.1  &  $2.42 \pm 0.16$  &  $74.5 \pm 5.0$     &  $51.5 \pm 3.4$     &  $61.4 \pm 4.1$     & $143.9 \pm 9.6$  & $102.2 \pm 6.8$ &  $891 \pm 59$
\enddata
\tablenotetext{a}{Pixel size: $0.768''$.}
\tablenotetext{b}{Within $6\times6$ pixels, corresponding to a $4.6''\times4.6''$ region.}
\tablenotetext{c}{Main peak: $09^{h}55^{m}51.28^{s}$, $+69^{\circ}40'45.5''$}
\tablenotetext{d}{Secondary peak: $09^{h}55^{m}52.68^{s}$, $+69^{\circ}40'48.5''$}
\tablenotetext{e}{Western Ridge: $09^{h}55^{m}50.47^{s}$, $+69^{\circ}40'43.9''$}
\tablenotetext{f}{Within $9\times9$ pixels, corresponding to a $6.8''\times6.8''$ region.}
\tablenotetext{g}{Within $50''\times75''$ ($65\times98$ pixel) region around center of M82}
\end{deluxetable}

\vfill

\clearpage

\vfill

\clearpage

\vfill

\clearpage

\vfill

\clearpage

\vfill


\begin{thebibliography}{33}
\expandafter\ifx\csname natexlab\endcsname\relax\def\natexlab#1{#1}\fi

\bibitem[{{Achtermann} \& {Lacy}(1995)}]{Achtermann.Lacy.1995}
{Achtermann}, J.~M., \& {Lacy}, J.~H. 1995, \apj, 439, 163

\bibitem[{{Adams} {et~al.}(2010){Adams}, {Herter}, {Gull}, {Schoenwald},
  {Henderson}, {Keller}, {De Buizer}, {Stacey}, \&
  {Nikola}}]{Adams.Herter.Gull.2010}
{Adams}, J.~D., {et~al.} 2010, in Society of Photo-Optical Instrumentation
  Engineers (SPIE) Conference Series, Vol. 7735, Society of Photo-Optical
  Instrumentation Engineers (SPIE) Conference Series

\bibitem[{{Appleton} {et~al.}(1981){Appleton}, {Davies}, \&
  {Stephenson}}]{Appleton.Davies.Stephenson.1981}
{Appleton}, P.~N., {Davies}, R.~D., \& {Stephenson}, R.~J. 1981, \mnras, 195,
  327

\bibitem[{{Beir{\~a}o} {et~al.}(2008){Beir{\~a}o}, {Brandl}, {Appleton},
  {Groves}, {Armus}, {F{\"o}rster Schreiber}, {Smith}, {Charmandaris}, \&
  {Houck}}]{Beirao.Brandl.Appleton.2008}
{Beir{\~a}o}, P., {et~al.} 2008, \apj, 676, 304

\bibitem[{{Dalcanton} {et~al.}(2009){Dalcanton}, {Williams}, {Seth}, {Dolphin},
  {Holtzman}, {Rosema}, {Skillman}, {Cole}, {Girardi}, {Gogarten},
  {Karachentsev}, {Olsen}, {Weisz}, {Christensen}, {Freeman}, {Gilbert},
  {Gallart}, {Harris}, {Hodge}, {de Jong}, {Karachentseva}, {Mateo}, {Stetson},
  {Tavarez}, {Zaritsky}, {Governato}, \&
  {Quinn}}]{Dalcanton.Williams.Seth.2009}
{Dalcanton}, J.~J., {et~al.} 2009, \apjs, 183, 67

\bibitem[{{de Grijs}(2001)}]{deGrijs.2001}
{de Grijs}, R. 2001, Astronomy and Geophysics, 42, 040000

\bibitem[{{de Mello} {et~al.}(2008){de Mello}, {Smith}, {Sabbi}, {Gallagher},
  {Mountain}, \& {Harbeck}}]{deMello.Smith.Sabbi.2008}
{de Mello}, D.~F., {Smith}, L.~J., {Sabbi}, E., {Gallagher}, J.~S., {Mountain},
  M., \& {Harbeck}, D.~R. 2008, \aj, 135, 548

\bibitem[{{Draine}(2003)}]{Draine.2003}
{Draine}, B.~T. 2003, \araa, 41, 241

\bibitem[{{F{\"o}rster Schreiber} {et~al.}(2001){F{\"o}rster Schreiber},
  {Genzel}, {Lutz}, {Kunze}, \&
  {Sternberg}}]{FoersterSchreiber.Genzel.Lutz.2001}
{F{\"o}rster Schreiber}, N.~M., {Genzel}, R., {Lutz}, D., {Kunze}, D., \&
  {Sternberg}, A. 2001, \apj, 552, 544

\bibitem[{{Gandhi} {et~al.}(2011){Gandhi}, {Isobe}, {Birkinshaw}, {Worrall},
  {Sakon}, {Iwasawa}, \& {Bamba}}]{Gandhi.Isobe.Birkinshaw.2011}
{Gandhi}, P., {Isobe}, N., {Birkinshaw}, M., {Worrall}, D.~M., {Sakon}, I.,
  {Iwasawa}, K., \& {Bamba}, A. 2011, \pasj, 63, 505

\bibitem[{{Greve} {et~al.}(2002){Greve}, {Wills}, {Neininger}, \&
  {Pedlar}}]{Greve.Wills.Neininger.2002}
{Greve}, A., {Wills}, K.~A., {Neininger}, N., \& {Pedlar}, A. 2002, \aap, 383,
  56

\bibitem[{{Herter} {et~al.}(2012{\natexlab{a}}){Herter}, {Vacca}, \&
  D.}]{Herter.Vacca.Adams.2012}
{Herter}, T.~L., {Vacca}, W.~D., \& D., A.~J. 2012{\natexlab{a}}, in prep.

\bibitem[{{Herter} {et~al.}(2012{\natexlab{b}}){Herter}, D., {De Buizer},
  {Gull}, {Schoenwald}, {Henderson}, {Keller}, {Nikola}, {Stacey}, \&
  {Vacca}}]{Herter.Adams.deBuizer.2012}
{Herter}, T.~L., {et~al.} 2012{\natexlab{b}}, \apjl, this volume

\bibitem[{{Keto} {et~al.}(2005){Keto}, {Ho}, \& {Lo}}]{Keto.Ho.Lo.2005}
{Keto}, E., {Ho}, L.~C., \& {Lo}, K.-Y. 2005, \apj, 635, 1062

\bibitem[{{Li} \& {Draine}(2001)}]{Li.Draine.2001}
{Li}, A., \& {Draine}, B.~T. 2001, \apj, 554, 778

\bibitem[{{Lipscy} \& {Plavchan}(2004)}]{Lipscy.Plavchan.2004}
{Lipscy}, S.~J., \& {Plavchan}, P. 2004, \apj, 603, 82

\bibitem[{{Matsushita} {et~al.}(2005){Matsushita}, {Kawabe}, {Kohno},
  {Matsumoto}, {Tsuru}, \& {Vila-Vilar{\'o}}}]{Matsushita.Kawabe.Kohno.2005}
{Matsushita}, S., {Kawabe}, R., {Kohno}, K., {Matsumoto}, H., {Tsuru}, T.~G.,
  \& {Vila-Vilar{\'o}}, B. 2005, \apj, 618, 712

\bibitem[{{Matsushita} {et~al.}(2000){Matsushita}, {Kawabe}, {Matsumoto},
  {Tsuru}, {Kohno}, {Morita}, {Okumura}, \&
  {Vila-Vilar{\'o}}}]{Matsushita.Kawabe.Matsumotot.2000}
{Matsushita}, S., {Kawabe}, R., {Matsumoto}, H., {Tsuru}, T.~G., {Kohno}, K.,
  {Morita}, K.-I., {Okumura}, S.~K., \& {Vila-Vilar{\'o}}, B. 2000, \apjl, 545,
  L107

\bibitem[{{Mayya} {et~al.}(2006){Mayya}, {Bressan}, {Carrasco}, \&
  {Hernandez-Martinez}}]{Mayya.Bressan.Carrasco.2006}
{Mayya}, Y.~D., {Bressan}, A., {Carrasco}, L., \& {Hernandez-Martinez}, L.
  2006, \apj, 649, 172

\bibitem[{{Panuzzo} {et~al.}(2010){Panuzzo}, {Rangwala}, {Rykala}, {Isaak},
  {Glenn}, {Wilson}, {Auld}, {Baes}, {Barlow}, {Bendo}, {Bock}, {Boselli},
  {Bradford}, {Buat}, {Castro-Rodr{\'{\i}}guez}, {Chanial}, {Charlot},
  {Ciesla}, {Clements}, {Cooray}, {Cormier}, {Cortese}, {Davies}, {Dwek},
  {Eales}, {Elbaz}, {Fulton}, {Galametz}, {Galliano}, {Gear}, {Gomez},
  {Griffin}, {Hony}, {Levenson}, {Lu}, {Madden}, {O'Halloran}, {Okumura},
  {Oliver}, {Page}, {Papageorgiou}, {Parkin}, {P{\'e}rez-Fournon}, {Pohlen},
  {Polehampton}, {Rigby}, {Roussel}, {Sacchi}, {Sauvage}, {Schulz}, {Schirm},
  {Smith}, {Spinoglio}, {Stevens}, {Srinivasan}, {Symeonidis}, {Swinyard},
  {Trichas}, {Vaccari}, {Vigroux}, {Wozniak}, {Wright}, \&
  {Zeilinger}}]{Panuzzo.Rangwala.Rykala.2010}
{Panuzzo}, P., {et~al.} 2010, \aap, 518, L37+

\bibitem[{{Rieke} {et~al.}(1980){Rieke}, {Lebofsky}, {Thompson}, {Low}, \&
  {Tokunaga}}]{Rieke.Lebofsky.Thompson.1980}
{Rieke}, G.~H., {Lebofsky}, M.~J., {Thompson}, R.~I., {Low}, F.~J., \&
  {Tokunaga}, A.~T. 1980, \apj, 238, 24

\bibitem[{{Sanders} {et~al.}(1991){Sanders}, {Scoville}, \&
  {Soifer}}]{Sanders.Scoville.Soifer.1991}
{Sanders}, D.~B., {Scoville}, N.~Z., \& {Soifer}, B.~T. 1991, \apj, 370, 158

\bibitem[{{Satyapal} {et~al.}(1997){Satyapal}, {Watson}, {Pipher}, {Forrest},
  {Greenhouse}, {Smith}, {Fischer}, \&
  {Woodward}}]{Satyapal.Watson.Pipher.1997}
{Satyapal}, S., {Watson}, D.~M., {Pipher}, J.~L., {Forrest}, W.~J.,
  {Greenhouse}, M.~A., {Smith}, H.~A., {Fischer}, J., \& {Woodward}, C.~E.
  1997, \apj, 483, 148

\bibitem[{{Satyapal} {et~al.}(1995){Satyapal}, {Watson}, {Pipher}, {Forrest},
  {Coppenbarger}, {Raines}, {Libonate}, {Piche}, {Greenhouse}, {Smith},
  {Thompson}, {Fischer}, {Woodward}, \& {Hodge}}]{Satyapal.Watson.Pipher.1995}
{Satyapal}, S., {et~al.} 1995, \apj, 448, 611

\bibitem[{{Siebenmorgen} \& {Kr{\"u}gel}(2007)}]{Siebenmorgen.Kruegel.2007}
{Siebenmorgen}, R., \& {Kr{\"u}gel}, E. 2007, \aap, 461, 445

\bibitem[{{Sun} {et~al.}(2005){Sun}, {Zhou}, {Chen}, {Burstein}, {Windhorst},
  {Ma}, {Byun}, {Jiang}, \& {Chen}}]{Sun.Zhou.Chen.2005}
{Sun}, W.-H., {et~al.} 2005, \apjl, 630, L133

\bibitem[{{Telesco} {et~al.}(1991){Telesco}, {Joy}, {Dietz}, {Decher}, \&
  {Campins}}]{Telesco.Joy.Dietz.1991}
{Telesco}, C.~M., {Joy}, M., {Dietz}, K., {Decher}, R., \& {Campins}, H. 1991,
  \apj, 369, 135

\bibitem[{{Weingartner} \& {Draine}(2001)}]{Weingartner.Draine.2001a}
{Weingartner}, J.~C., \& {Draine}, B.~T. 2001, \apj, 548, 296

\bibitem[{{Wei{\ss}} {et~al.}(2001){Wei{\ss}}, {Neininger}, {H{\"u}ttemeister}, 
   \& {Klein}}]{Weiss.Neininger.Huettenmeister.2001}
{Wei{\ss}}, A., {Neininger}, N., {H{\"u}ttemeister}, S., \& {Klein}, U. 2001,
   \aap, 365, 571

\bibitem[{{Wills} {et~al.}(2000){Wills}, {Das}, {Pedlar}, {Muxlow}, \&
  {Robinson}}]{Wills.Das.Pedlar.2000}
{Wills}, K.~A., {Das}, M., {Pedlar}, A., {Muxlow}, T.~W.~B., \& {Robinson},
  T.~G. 2000, \mnras, 316, 33

\bibitem[{{Young} {et~al.}(2012){Young}, {Herter}, {De Buizer}, D., {Becklin},
  {Gehrz}, {Gull}, {Harvey}, {Helton}, {Henderson}, {Keller}, {Krabbe},
  {Marcum}, {Megeath}, {Morris}, {Nikola}, {Reach}, {Roellig}, {Sandell},
  {Sankrit}, {Schoenwald}, Y., {Stacey}, {Temi}, {Tielens}, {Vacca}, \&
  {Zinnecker}}]{Young.Herter.deBuizer.2012}
{Young}, E.~T., {et~al.} 2012, \apjl, this volume

\bibitem[{{Yun} {et~al.}(1994){Yun}, {Ho}, \& {Lo}}]{Yun.Ho.Lo.1994}
{Yun}, M.~S., {Ho}, P.~T.~P., \& {Lo}, K.~Y. 1994, \nat, 372, 530

\end{thebibliography}
\end{document}